\documentclass[preprint,
               sort&compress,
              ]{elsarticle}

\usepackage{amsmath}
\usepackage{amssymb}
\usepackage{mathtools}
\usepackage{algorithm}
\usepackage{algorithmicx}
\usepackage{algpseudocode}
\usepackage{hyperref}
\usepackage[normalem]{ulem}
\usepackage{xcolor}

\renewcommand{\vec}[1]{{\ensuremath{\mathchoice
                     {\mbox{\boldmath$\displaystyle\mathbf{#1}$}}
                     {\mbox{\boldmath$\textstyle\mathbf{#1}$}}
                     {\mbox{\boldmath$\scriptstyle\mathbf{#1}$}}
                     {\mbox{\boldmath$\scriptscriptstyle\mathbf{#1}$}}}}}%

\newcommand{\tens}[1]{{\ensuremath{\mathchoice
                     {\mbox{$\displaystyle\mathsf{#1}$}}
                     {\mbox{$\textstyle\mathsf{#1}$}}
                     {\mbox{$\scriptstyle\mathsf{#1}$}}
                     {\mbox{$\scriptscriptstyle\mathsf{#1}$}}}}}%

\renewcommand{\d}{\ensuremath{\partial}}

\newcommand{\eq}{\ensuremath{\text{eq}}}
\newcommand{\nq}{\ensuremath{\text{neq}}}
\newcommand{\Fvec}{\ensuremath{\vec{F}}}

\newcommand{\Vvec}{\ensuremath{\vec{V}}}
\newcommand{\Xvec}{\ensuremath{\vec{X}}}
\newcommand{\avec}{\ensuremath{\vec{a}}}
\newcommand{\cvec}{\ensuremath{\vec{c}}}
\newcommand{\fvec}{\ensuremath{\vec{f}}}

\newcommand{\rvec}{\ensuremath{\vec{r}}}
\newcommand{\uvec}{\ensuremath{\vec{u}}}
\newcommand{\vvec}{\ensuremath{\vec{v}}}
\newcommand{\xvec}{\ensuremath{\vec{x}}}
\newcommand{\zetavec}{\ensuremath{\vec{\zeta}}}
\newcommand{\Pitens}{\ensuremath{\tens{\Pi}}}

\newcommand{\1}{\ensuremath{\mathsf{I}}}
\newcommand{\Iop}{\ensuremath{\mathcal{I}}}
\newcommand{\Lambdaop}{\ensuremath{\mathsf{\Lambda}}}
\newcommand{\Omegaop}{\ensuremath{\mathsf{\Omega}}}

\begin{document}


\title{A unified operator splitting approach for multi-scale
fluid-particle coupling in the lattice Boltzmann method}

\author{Ulf D. Schiller}\fnref{fn1}

\ead{u.schiller@ucl.ac.uk}

\address{%
  Theoretical Soft Matter and Biophysics,
  Institute of Complex Systems,
  Forschungszentrum~J\"ulich,
  52425 J\"ulich,
  Germany
}

\fntext[fn1]{Present address: Centre for Computational Science, University College London, 20 Gordon Street, London WC1H 0AJ, United Kingdom}


\date{\today}

\begin{abstract}
A unified framework to derive discrete time-marching schemes for
coupling of immersed solid and elastic objects to the lattice
Boltzmann method is presented. Based on operator splitting for the
discrete Boltzmann equation, second-order time-accurate schemes for
the immersed boundary method, viscous force coupling and external
boundary force are derived. Furthermore, a modified formulation of the
external boundary force is introduced that leads to a more accurate
no-slip boundary condition. The derivation also reveals that the
coupling methods can be cast into a unified form, and that the
immersed boundary method can be interpreted as the limit of force
coupling for vanishing particle mass. In practice, the ratio between
fluid and particle mass determines the strength of the force transfer
in the coupling. The integration schemes formally improve the accuracy
of first-order algorithms that are commonly employed when coupling
immersed objects to a lattice Boltzmann fluid. It is anticipated that
they will also lead to superior long-time stability in simulations of
complex fluids with multiple scales.
%
\end{abstract}


\maketitle


\section{Introduction}

The lattice Boltzmann method (LBM) \cite{Succi-LB} is a popular
approach to simulate hydrodynamic phenomena and is widely used as an
alternative to continuum-based approaches to solving the Navier-Stokes
equations. The LBM is a mesoscopic model that bridges between kinetic
theory and macroscopic hydrodynamics, i.e., it reproduces the
conservation laws of the Boltzmann equation (in particular mass and
momentum conservation) on macroscopic time and length scales
\cite{He1997a,He1997b,Shan1998,Junk2005a}.
Since the LBM enables simple and accurate incorporation of geometric
boundaries, it is widely applied to fluid-structure interaction
problems \cite{Krafczyk2001,Geller2006,DeRosis2013}. 
%

In recent years, there has been growing interest in using the lattice
Boltzmann method to simulate momentum transport and hydrodynamic
interactions in complex fluids and soft matter systems
\cite{Duenweg2008,Aidun2010}.  Such systems concern objects that are
immersed in a continuum solvent, for example, colloidal suspensions
\cite{Ladd1994a,Ladd2001}, polymer solutions
\cite{Ahlrichs1999,Pham2009,Ladd2009}, or biological cells
\cite{Alexeev2005,Dupin2007,Wu2010}. These objects can be represented
by a set of ``coarse-grained'' particles which interact via effective
potentials and reproduce the correct dynamics on physically
interesting scales. Since these scales are typically large compared to
atomistic scales, the solvent can be treated as a continuum and the
hydrodynamic fields represent slow degrees of freedom of the
microscopic kinetics.
The standard lattice Boltzmann algorithm gives a second-order accurate
approximation of the macroscopic transport equations for the conserved
moments, however, for grid-scale Reynolds numbers above unity the
non-conserved moments may be subject to oscillations that can cause
nonlinear instabilities \cite{Brownlee2007}.
%
Soft matter flows are usually in the creeping flow regime, i.e., they
are incompressible (low Mach number) and inertia can be neglected (low
Reynolds number). 
%

%
%
%

The dynamics of complex fluids with immersed objects can be strongly
influenced by the coupling between solute and solvent degrees of
freedom. The motion of a solute particle creates a perturbation of the
solvent that spreads out to other particles. Hence, the particle
movements become correlated due to viscous momentum diffusion. The
idea of hybrid fluid-particle methods is to simulate the momentum
transport in an explicit solvent with the LBM while the solute objects
are coupled to the flow field by some sort of interaction force
\cite{Duenweg2008}.
If the immersed objects have internal degrees of freedom,
e.g. polymers, membranes or cells, the equations of motion for these
degrees of freedom have to be solved simultaneously with the fluid
dynamics. Various approaches have been proposed to solve this task,
such as the immersed boundary method (IBM)~\cite{Peskin2002}, force
coupling through viscous drag~\cite{Ahlrichs1999}, or external
boundary force (EBF) \cite{Wu2010}. They have in common that a set of
Lagrangian particles or marker points is added to the lattice
Boltzmann fluid, which require the simultaneous integration of
Newton's equations of motion, e.g., by molecular dynamics (MD)
techniques.

Integration schemes for MD can be faced with spurious effects due to
the time discretization, for example, systematic drift of energy or
temperature, artificial spatial correlations between particles
\cite{Serrano2006}, and numerical instabilities \cite{Schlick1998}.  A
number of improved algorithms has been introduced,
e.g. \cite{Tuckerman1992,Ricci2003,Serrano2006,Bussi2007}, which are
commonly based on a Trotter expansion of the Liouville propagator
\cite{Trotter1959,Kato1978,Suzuki1985}. Integration schemes derived by
splitting methods can improve accuracy and lead to symplectic
algorithms that are usually also more stable
\cite{Tuckerman1992,McLachlan2002,Bou-Rabee2013}.

In this communication, a general approach to derive time-marching
schemes to solve hybrid fluid-particle systems is presented. It is
based on the interpretation of the lattice Boltzmann equation as a
Strang splitting scheme \cite{Dellar2013} combined with common
operator splitting techniques used in molecular dynamics (MD) and
related methods
\cite{FrenkelSmit,Tuckerman1992,Skeel1997,Shardlow2003,
DeFabritiis2006,Bou-Rabee2013}. The operator splitting makes it
straightforward to derive explicit time-discretizations, in
particular, second-order time-accurate integration schemes for the
various coupling methods are proposed. Furthermore, a modified
external boundary force is proposed that ensures the no-slip boundary
condition after application of the force operator. The latter also
reveals that the IBM can be interpreted as the limit of force coupling
methods for vanishing particle mass, and that the force transfer
during coupling is controlled by the ratio between fluid and particle
mass. This notion may also be applicable to other mesoscopic methods
such as multi-particle collision dynamics
\cite{Malevanets1999,Gompper2009}.

The remainder of this work is organized as follows: In section
\ref{sec:Boltzmann} and \ref{sec:splitting}, the derivation of the
lattice Boltzmann equation and its interpretation as an operator
splitting scheme \cite{Dellar2013} are briefly reviewed. In section
\ref{sec:CSC-LB}, an alternative lattice Boltzmann algorithm in the
form of a collide-stream-collide scheme is outlined and analyzed. The
operator splitting is then extended to several coupling methods, i.e.,
the immersed boundary method (section \ref{sec:IBM}), force coupling
(section \ref{sec:force-coupling}), and external boundary force
(section \ref{sec:ebf}). Finally, multiple time-step schemes are
briefly discussed in section~\ref{sec:multi-timestep}.

\section{Discrete Boltzmann equation}
\label{sec:Boltzmann}

In the kinetic theory of gases, the dynamics and transport properties
of fluids are determined by the Boltzmann equation
\cite{Cercignani1988}
%
\begin{equation}\label{eq:Boltzmann}
\left(\frac{\d}{\d t} + \cvec\cdot\frac{\d}{\d\xvec} +
  \avec \cdot\frac{\d}{\d\cvec}\right) f(\xvec,\cvec,t) =
- \Omegaop ( f - f^\eq ) . 
\end{equation}
Here, $f(\xvec,\cvec,t)$ is the one-particle distribution function at
position $\xvec$, velocity $\vvec$, and time $t$, and the collision
operator $\Omegaop$ on the right-hand side has been linearized around
the equilibrium distribution $f^\eq$ \cite{Bhatnagar1954,Gross1959}.
In order to discretize the Boltzmann equation \eqref{eq:Boltzmann},
the moment integrals of the distribution function are formally
evaluated by a Gauss-Hermite quadrature, where the abscissae of the
quadrature comprise a set of discrete velocities $\cvec_i$
\cite{Shan1998,Shan2006}. This procedure leads to the {discrete
Boltzmann equation}, which is also the basis of discrete velocity
models \cite{Broadwell1964a,Luo2000}
%
\begin{equation}\label{eq:discrete-Boltzmann}
\left( \frac{\d}{\d t} + \cvec_i \cdot \frac{\d}{\d \xvec} \right) f_i = - \sum_j \Omega_{ij} \left( f_j - f_j^{\eq} \right) + G_i ,
\end{equation}
%
where $G_i$ is a body force corresponding to the acceleration term
$-\avec\cdot\nabla_\cvec f$. 
The standard approach to space-time discretization of the discrete
Boltzmann equation is integrating along the characteristic $(\xvec+h
\cvec_i, t+h)$ and approximating the integral of the collision term by
the trapezium rule \cite{He1997a,He1997b}. This leads to a non-linear
implicit equation for $f_i(\xvec+h \cvec_i, t+h)$ that can be turned
into an explicit scheme by an appropriate re-definition of variables
\cite{He1998}. This explicit scheme is the lattice Boltzmann equation
in the standard form
\cite{Benzi1992,Chen1998,He1998a,Guo2002}
\begin{equation}\label{eq:LBE}
\tilde f_i(\xvec+h\cvec_i,t+h) = \tilde f_i(\xvec,t) - \sum_j
\Lambda_{ij} \left( \tilde f_j - f_j^{\eq} \right) + \sum_j (\delta_{ij} - \frac{1}{2} \Lambda_{ij}) h G_j(t) ,
\end{equation}
where the new variables are \cite{He1998,Dellar2003}
\begin{equation}\label{eq:He-trafo1}
\tilde f_i(t) = f_i(t) + \sum_j \frac{h}{2} \Omega_{ij} \left( f_j - f_j^{\eq} \right) - \frac{h}{2} G_i(t) .
\end{equation}
It is important to make the distinction between $f_i$ and the new
variables $\tilde f_i$ and to replace the collision matrix $h\Omegaop$
by the discrete collision matrix $\Lambdaop =
(1+\frac{h}{2}\Omegaop)^{-1}h\Omegaop$
\cite{Henon1987,Xu2003,Dellar2003}. Without these replacements,
Eq.~\eqref{eq:LBE} is only a first-order accurate discretization of
the Boltzmann equation \cite{Ohwada1998}. While $\Lambdaop$ enters in
the collision step and leads to the correct viscosity, the variable
transformation \eqref{eq:He-trafo1} is essential to evaluate the
macroscopic flow fields which have to be obtained as moments of the
back-transformed variables $f_i$:
%
%
%
\begin{align}
\rho &= \sum_i f_i ,
& \rho \uvec &= \sum_i f_i \cvec_i ,
& \Pitens &= \sum_i f_i \cvec_i \cvec_i .
\end{align}
In the absence of forces, the back-transformation is only relevant for
evaluating the (non-equilibrium) stress and does not change the
mass and momentum densities.

Dellar \cite{Dellar2013} has recently pointed out that the lattice
Boltzmann equation can also be interpreted as a Strang splitting
scheme. In the following, operator splitting for the discrete
Boltzmann equation is revisited, which will serve as the basis to
derive accurate coupling schemes in later sections.  The operator
splitting approach will also provide a complementary view on the role
of Eq.~\eqref{eq:He-trafo1}.

\section{Operator splitting for the discrete Boltzmann equation}
\label{sec:splitting}

In the operator splitting interpretation put forward by
Dellar \cite{Dellar2013}, the lattice Boltzmann equation is split into
decoupled streaming, collision and forcing steps. The idea is to find
approximate solutions to the individual steps and combine them into a
time-marching scheme, the accuracy of which depends on the order of
the applied splitting and the accuracy of the individual approximate
solutions.
Splitting methods \cite{McLachlan2002} make use of decomposition
formulas for exponential operators where the exponential of a sum of
non-commuting operators is approximated by a symmetrized product of
exponentials of the individual operators
\cite{Trotter1959,Kato1978,Suzuki1985}. For the simplest case of two
operators, the resulting splitting scheme is equivalent to the
difference scheme introduced by Strang \cite{Strang1968}.  
%
%
%
The decomposition formulas are valid for linear operators and can be
generalized to nonlinear operators by application of the
Baker-Campbell-Hausdorff formula
\cite{Groebner1960,Varadarajan1974,Hairer2006}. The truncated Taylor
series of the exponentials can be applied to sufficiently
differentiable functions and yield formally accurate splitting schemes
\cite{Yoshida1990,Hairer2006}.
Hence it is justified to use splitting schemes for the Boltzmann
kinetic equation \cite{Bobylev2001} although the collision operator
depends non-linearly on the flow velocity through the equilibrium
distribution. The remainder of this section closely follows the work
of Dellar \cite{Dellar2013}. However, instead of combining adjacent
parts of different time-steps to recover the standard lattice
Boltzmann algorithm up to a transformation of variables, we will
combine the sub-parts of one time-step into an alternative update
scheme in section~\ref{sec:CSC-LB}. This provides a complementary view
on the splitting scheme and clarifies the role of the transformed
variables as intermediate values.

The discrete Boltzmann equation \eqref{eq:discrete-Boltzmann} can be
written in terms of operators as
\begin{equation}\label{eq:split-lb}
\frac{\d}{\d t} f_i(\xvec,t) = \left( \mathcal{S} + \mathcal{C} + \mathcal{F} \right) f_i(\xvec,t) ,
\end{equation}
where the operators for streaming $\mathcal{S}$, collisions
$\mathcal{C}$, and forces $\mathcal{F}$ are
\begin{subequations}\label{eq:decoupled}
\begin{align}
\mathcal{S} f_i &=  - \cvec_i \cdot \frac{\d}{\d \xvec} f_i , \\
\mathcal{C} f_i &= - \sum_j \Omega_{ij} \left( f_j - f_j^{\eq} \right) , \\
\mathcal{F} f_i &= G_i .
\end{align}
\end{subequations}
The action of forces is usually incorporated into the collision
operator $\mathcal{C}$, but it can also be treated as a third
decoupled step \cite{Salmon1999a,Salmon1999b,Dellar2013}. Here, we
adopt the splitting using three decoupled parts for streaming,
collisions and forces. The motivation is that the force operator
couples the interactions with the lattice Boltzmann equation and may
be implemented more cleanly if treated separately.

The formal solutions of the decoupled equations \eqref{eq:decoupled}
are
\begin{subequations}
\begin{align}
f_i(\xvec,t+h) &= e^{h\mathcal{S}} f_i(\xvec,t) \approx \mathsf{S} f_i(\xvec,t) , \label{eq:streaming} \\
f_i(\xvec,t+h) &= e^{h\mathcal{C}} f_i(\xvec,t) \approx \mathsf{C} f_i(\xvec,t) , \label{eq:collision} \\
f_i(\xvec,t+h) &= e^{h\mathcal{F}} f_i(\xvec,t) \approx \mathsf{F} f_i(\xvec,t) ,
\label{eq:force}
\end{align}
\end{subequations}
where the three operators $\mathsf{S} \approx \exp[h\mathcal{S}]$,
$\mathsf{C} \approx \exp[h\mathcal{C}]$, and $\mathsf{F} \approx
\exp[h\mathcal{F}]$ are introduced as approximations to the separate
exponential operators. Note that the index $i$ and the coordinate
$\xvec$ in the streaming part \eqref{eq:streaming} are to be
understood as ranging over all possible values because streaming is a
non-local operation. In this sense, the operator $\exp[h\mathcal{S}]$
acts on the whole ensemble of distributions.
The discrete approximations for the decoupled steps are obtained
below.

The formal solution to \eqref{eq:split-lb} is given by
\begin{equation}\label{eq:LB-evolution}
f_i(\xvec,t+h) = e^{h(\mathcal{S}+\mathcal{C}+\mathcal{F})}
f_{i}(\xvec,t) ,
\end{equation}
where again the index $i$ and the coordinates $\xvec$ are ranging over
all values.
A second-order Trotter-Suzuki decomposition
\cite{Trotter1959,Suzuki1985} of the exponential operator leads to the
splitting scheme
\begin{equation}\label{eq:LB-splitting}
\begin{split}
e^ { h ( \mathcal{S} + \mathcal{C} + \mathcal{F} ) } &= e^ {\frac{h}{2} \mathcal{F} } e^ { \frac{h}{2} \mathcal{C} } e^{ h \mathcal{S} } e^{ \frac{h}{2} \mathcal{C} } e^{ \frac{h}{2} \mathcal{F} } + O(h^3) , \\
&= \mathsf{F}^{\frac{1}{2}} \mathsf{C}^{\frac{1}{2}} \mathsf{S} \mathsf{C}^{\frac{1}{2}} \mathsf{F}^{\frac{1}{2}} + O(h^3) ,
\end{split}
\end{equation}
where the operators $\mathsf{S}$, $\mathsf{C}^\frac12$, and
$\mathsf{F}^\frac12$ are the discrete approximations of the
exponential operators for a step $h$ and $h/2$, respectively.
The lattice Boltzmann algorithm requires that the streaming part
$\mathsf{S}$ must not be split in order to be compatible with the
lattice structure. For second-order accuracy, $\mathsf{F}^\frac{1}{2}$
and $\mathsf{C}^\frac{1}{2}$ have to be arranged symmetrically but
their order is in principle arbitrary. The choice to put
$\mathsf{F}^\frac{1}{2}$ to the outside is motivated by the coupling
algorithms discussed below. For small enough forces
$\mathsf{F}^\frac{1}{2}$ and $\mathsf{C}^\frac{1}{2}$ commute,
cf. section~\ref{sec:forces}.

\subsection{Streaming}

The streaming part \eqref{eq:streaming} can be written as a derivative along the
characteristic $(x+h'\cvec_i,t+h')$
\begin{equation}
\left( \frac{\d}{\d t} + \cvec_i\cdot\frac{\d}{\d \xvec} \right) f_i(\xvec,t) = \frac{d}{d h'} f_i(\xvec+h'\cvec_i,t+h') = 0 ,
\end{equation}
and can be integrated from $0$ to $h$ to give
\begin{equation}
\begin{gathered}
\int_0^h \frac{d}{dh'} f_i(\xvec+h'\cvec_i,t+h') dh' = f_i(\xvec+h\cvec_i,t+h) - f_i(\xvec,t) = 0 .
\end{gathered}
\end{equation}
We thus get the discrete streaming step
\begin{equation}
\begin{gathered}
f_i(\xvec+h\cvec_i,t+h) = f_i(\xvec,t) ,
\end{gathered}
\end{equation}
from which the spatial grid structure emerges, i.e., only coordinates
connected by the discrete vectors $h \cvec_i$ are admissible. 


\subsection{Collisions}


The collision part \eqref{eq:collision} leaves the equilibrium
distribution invariant since $\mathcal{C} f_i^\eq = 0$. Hence we
consider only the non-equilibrium part $f_i^\nq = f_i - f_i^\eq$ and
the differential equation
\begin{equation}
\frac{d}{d t} \fvec^\nq = \mathcal{C} \fvec^\nq = - \Omegaop \, \fvec^\nq ,
\end{equation}
where we have introduced the vector notation
$\fvec=(f_0,f_1,\dots,f_q)^T$.
An approximate solution is obtained using the $O(h^3)$ accurate
Crank-Nicolson rule \mbox{\cite{Dellar2003,Dellar2013}}
\begin{equation}\label{eq:CN-approx}
\begin{split}
  \fvec^\nq(t+h)-\fvec^\nq(t) &= - \frac{h}{2} \Omegaop \left[ \fvec^\nq(t+h) + \fvec^\nq(t) \right] , \\
\end{split}
\end{equation}
which leads to
\begin{equation}\label{eq:CN-coll}
\begin{split}
  \fvec^\nq(t+h) = \left( \1 + \frac{h}{2} \Omegaop \right)^{-1} 
  \left( \1 - \frac{h}{2} \Omegaop \right) \fvec^\nq(t) .
\end{split}
\end{equation}
A comparison with the exact solution 
\begin{equation}\label{eq:exact-coll}
  \fvec^\nq(t+h) = \exp [ - h \Omegaop ] \, \fvec^\nq(t)
\end{equation}
shows that the accuracy of \eqref{eq:CN-coll} is in fact
$O((h/\tau)^3)$, which can cause nonlinear instabilities for grid
scale Reynolds number $h/\tau$ above unity \cite{Brownlee2007}.  On
the other hand, the analysis of Dellar \cite{Dellar2013} revealed that
the exact solution \eqref{eq:exact-coll} leads to excessively rapid
decay of shear modes as $\tau \lesssim h$, while the Crank-Nicolson
approximation \eqref{eq:CN-coll} gives accurate decay rates even for
small collision times. The reason for this somewhat counter-intuitive
result lies in the compensation of errors introduced by the splitting
of the Boltzmann equation into uncoupled steps on one hand, and the
Crank-Nicolson approximation of the collision operator on the other.


Equation \eqref{eq:CN-coll} can be rewritten as
\begin{equation}
\begin{split}
\mathsf{C} \fvec^\nq &= \left[ \1 - \left( \1 + \frac{h}{2} \Omegaop \right)^{-1} h \Omegaop \right] \, \fvec^\nq \\
&= \left( \1 - \Lambdaop \right) \, \fvec^\nq
\end{split}
\end{equation}
with the discrete collision matrix \cite{Dellar2003}
\begin{equation}\label{eq:collision-operator}
\Lambdaop = \left( \1 + \frac{h}{2}\Omegaop \right)^{-1}
h \Omegaop .
\end{equation}
This corresponds to the familiar replacement of the collision time
$\tau$ by $\tau+h/2$ in the discrete scheme \cite{Henon1987}.
The half-step collision update $\mathsf{C}^\frac12$ in the splitting
scheme \eqref{eq:LB-splitting} is obtained by replacing $h$ by $h/2$
in Eq.~\eqref{eq:collision-operator}
\begin{equation}\label{eq:collision-operator2}
\Lambdaop_\frac12 = \left( \1 + \frac{h}{4}\Omegaop \right)^{-1}
\frac{h}{2} \Omegaop .
\end{equation}

We can convince ourselves that with $\Lambdaop$ and
$\Lambdaop_\frac12$ as in \eqref{eq:collision-operator} and
\eqref{eq:collision-operator2}, two applications of the half-step
collisions $\mathsf{C}^\frac12$ are up to order $O(h^3)$ equivalent to
one full-step collision~$\mathsf{C}$
%
\begin{equation}
\begin{split}
\mathsf{C}^\frac12\mathsf{C}^\frac12 \fvec^\nq
&= (\1-\Lambdaop_\frac12)^2 \, \fvec^\nq \\
&= (\1-\Lambdaop) \, \fvec^\nq + O(h^3) \\
&= \mathsf{C} \fvec^\nq + O(h^3) .
\end{split}
\end{equation}
The order of accuracy follows from $ (\1-\Lambdaop_\frac12)^2 = ( \1 -
\Lambdaop) + O(h^3) $.
%
%
Given $\Lambdaop$, we can also exactly define $\1-\Lambdaop_\frac12 =
\sqrt{\1-\Lambdaop}$, where the square-root is to be understood in the
sense of a Cholesky decomposition. For the eigenvalues, we simply have
$1-\lambda_\frac12 = \sqrt{1-\lambda}$.
In contrast, the approximation $\mathsf{C}^\frac12 = \frac12(\1 +
\mathsf{C})$ is only $O(h^2)$ accurate since $\Lambdaop_\frac12 =
\Lambdaop/2+O(h^2)$
%
\begin{equation}\label{eq:half-col-approx}
\begin{split}
\mathsf{C}^\frac12 \fvec^\nq
&= \left( \1 - \Lambdaop_\frac12 \right) \fvec^\nq \\
&= \left( \1 - \frac{1}{2} \Lambdaop \right) \fvec^\nq + O(h^2)
= \frac{1}{2} ( \1 + \mathsf{C} ) \fvec^\nq + O(h^2) .
\end{split}
\end{equation}
When the half-step collisions are applied in every iteration of the
time-marching scheme, the accuracy of $\mathsf{C}^\frac12$ is essential
to maintain the global order $O(h^2)$, cf. section \ref{sec:CSC-LB}.

For the sake of completeness, we mention that in the multi-relaxation
time approach (MRT), the collisions are typically implemented in the
space of moments $m_k$ of the $f_i$:
\begin{equation}\label{eq:mrt-coll}
\begin{split}
\mathsf{C} \, m_k^\nq 
    &= (1 - \lambda_k) \left[ m_k - m_k^\eq \right] .
\end{split}
\end{equation}
Here, $\lambda_k$ is the $k$-th eigenvalue of the
collision matrix which is diagonal in moment space. In the fluctuating
lattice Boltzmann method \cite{Ladd1994a,Adhikari2005,Duenweg2007}, a
noise term is added to the collision step
\begin{equation}\label{eq:mrt-fluct}
\mathsf{C} \, m_k^\nq = (1 - \lambda_k) \left[ m_k - m_k^\eq \right] + \sqrt{\rho} \varphi_k r_k ,
\end{equation}
where $r_k$ is a Gaussian random number, and the amplitude $\varphi_k
= \sqrt{\mu b_k \lambda_k^2}$ is related to the eigenvalue
$\lambda_k$ and the temperature through $\mu = a^{-3} k_b T /
c_s^2$. $b_k$ is the length of the $k$-th basis vector of the specific
lattice model. Equations \eqref{eq:mrt-coll} and \eqref{eq:mrt-fluct}
are the basis for the implementation of the collision step in
Algorithm~\ref{alg:csc}.  It is worth noting that the fluctuating
collision operator can be split into half-steps by a proper choice of
the eigenvalues and noise strength. A backwards half-step application
$\mathsf{C}^{-\frac12}$ as needed below in equation
\eqref{eq:LB-splitting-combined}, however, is not straightforward to
perform without distorting the correlations of the moments.

%

\subsection{Forces}
\label{sec:forces}

The force operator $\mathsf{F}$ in Eq.~\eqref{eq:force} can be
discretized using the midpoint rule
\begin{subequations}
\begin{equation}\label{eq:force-midpoint}
\begin{split}
\fvec(t+h) &= \mathsf{F} \fvec(t) = \fvec(t) + h \vec{G}(t+\frac{h}{2}) ,
\end{split}
\end{equation}
or alternatively with the Crank-Nicolson rule
\begin{equation}\label{eq:force-cn}
\begin{split}
\fvec(t+h) &= \mathsf{F} \fvec(t) = \fvec(t) + h \frac{\vec{G}(t+h)+\vec{G}(t)}{2}.
\end{split}
\end{equation}
\end{subequations}
The force term $\vec{G}=(G_1,G_2,\dots,G_q)^T$ has the same first
moments as the acceleration term $-\avec \cdot \nabla_\cvec f$
\cite{Luo1998,Guo2002}
\begin{align}\label{eq:force-moments}
\sum_i G_i &= 0, & \sum_i \cvec_i G_i &= \Fvec, & \sum_i \cvec_i \cvec_i G_i &= \Fvec \uvec + \uvec \Fvec,
\end{align}
where $\Fvec=\rho \avec$. Note that the minus sign for bringing $\avec
\cdot \nabla_\cvec f$ to the right hand side is compensated by another
minus sign that arises in the Hermite expansion of $\nabla_\cvec f$
\cite{Martys1998}.
%
%
In terms of the first moments, Eqs.~\eqref{eq:force-midpoint} and
\eqref{eq:force-cn} hence read
\begin{subequations}\label{eq:LB-force-moments}
\begin{align}
\begin{split}
\uvec(t+h)   &= \uvec(t) + \frac{h}{\rho} \Fvec(t+\frac{h}{2}) ,\\
\Pitens(t+h) &= 
\begin{multlined}[t]
\Pitens(t) + h \big[ \uvec(t+\frac{h}{2}) \Fvec(t+\frac{h}{2})  + \Fvec(t+\frac{h}{2}) \uvec(t+\frac{h}{2}) \big],
\end{multlined}
\end{split}
&& \text{(midpoint)} \\
\intertext{and}
\begin{split}
\uvec(t+h) &= \uvec(t) + \frac{h}{2\rho} \left( \Fvec(t+h) + \Fvec(t)
\right),\\
\Pitens(t+h) &= 
\begin{multlined}[t][.5\textwidth]
\Pitens(t) + \frac{h}{2} \big[ \uvec(t) \Fvec(t) +
\Fvec(t) \uvec(t) \\ + \uvec(t+h) \Fvec(t+h) + \Fvec(t+h) \uvec(t+h)
\big] .
\end{multlined}
\end{split}
&& \text{(Crank-Nicolson)}
\end{align}
\end{subequations}
If the force is constant during a time step $h$, Crank-Nicolson and
midpoint are both equivalent to forward Euler discretization, where
the midpoint velocity $\uvec(t+h/2)$ has to be used in the stress
update.
%


It is worthwhile to note that up to $O(\Fvec^2)$, the force operator
$\mathsf{F}$ only acts on the equilibrium distribution and leaves the
non-equilibrium part unchanged
\begin{equation}
\begin{split}
\fvec(t+h) - \fvec^\eq(\rho,\uvec(t+h)) &= \fvec(t) + h \vec{G}(t+\frac{h}{2}) - \fvec^\eq(\rho,\uvec(t+h)) \\
&= \fvec(t) - \fvec^\eq(\rho,\uvec(t)) + O(\Fvec^2) ,
\end{split}
\end{equation}
which can easily be checked using the truncated Hermite expansion of
the force term $G_i$ and the equilibrium distribution $f_i^\eq$
\cite{Luo1998,Martys1998,Guo2002}.
%
%
As a consequence, the force and collision operators approximately
commute
\begin{equation}
\begin{split}
\mathsf{C}\mathsf{F} \fvec(t) &= \mathsf{F}\fvec(t) - \Lambdaop \mathsf{F} \left[ \fvec(t) - \fvec^\eq(\rho,\uvec(t)) \right] \\
&= \fvec(t) + h \vec{G} - \Lambdaop \left[ \fvec(t) - \fvec^\eq(\rho,\uvec(t)) \right] + O(\Fvec^2) \\
&= \mathsf{C} \fvec(t) + h \vec{G} + O(\Fvec^2) \\
&= \mathsf{F}\mathsf{C} \fvec(t) + O(\Fvec^2) .
\end{split}
\end{equation}
For small enough forces the order of separate collisions and forcing
thus does not matter. Note that throughout this section we have
discussed a full time step. The action of the half-step force operator
$\mathsf{F}^\frac12$ is obtained by simply replacing $h$ by $h/2$. We
will see below that the splitting scheme leads to the familiar
transformation of the force term required for second-order accuracy
\cite{Guo2002,He1998a,Luo1998}.

\section{Collide-stream-collide lattice Boltzmann}
\label{sec:CSC-LB}

Strang or Trotter-Suzuki splitting as in Eq.~\eqref{eq:LB-splitting}
leads to a sequence of operations which involve the half-step collision
operator $\mathsf{C}^\frac12$. If it is the first and last operation,
then upon concatenation of time-steps the adjacent collision
half-steps can be combined, and the standard LB algorithm is recovered
\cite{Dellar2013}
\begin{equation}\label{eq:LB-splitting-combined}
\left[ \mathsf{C}^\frac12 \mathsf{S} \mathsf{C}^\frac12 \right] ^n = \
\mathsf{C}^\frac12 \left[ \mathsf{S} \mathsf{C} \right]^n
\mathsf{C}^{-\frac12} ,
\end{equation}
where the forces are included in the collision operator. At the
beginning and end of the simulation the half-step collisions
$\mathsf{C}^{-\frac12}$ and $\mathsf{C}^{\frac12}$ have to be applied,
respectively.
In this work, we group the operations such that the force operator
$\mathsf{F}$ does not interrupt the collision and streaming steps,
%
%
which facilitates the implementation of hybrid coupling algorithms with
sub-splitting of $\mathsf{F}$ (cf. section~\ref{sec:multi-timestep}).

Before developing the details of the coupling, we analyze the
splitting $\mathsf{C}^\frac12 \mathsf{S} \mathsf{C}^\frac12$ for the
LB part. The latter suggests a collide-stream-collide lattice
Boltzmann update with two half-step collisions before and after the
streaming. Let $f_i^*$ denote the populations after the first
collision step, and $\bar f_i$ the populations after the streaming
step. Then the three separate steps are
\begin{subequations}\label{eq:csc}
\begin{align}
&\fvec(\xvec,t) = 
\begin{multlined}[t]
\bar \fvec(\xvec,t) - \Lambdaop_{\frac12} \left[ \bar \fvec(\xvec,t) - \fvec^\eq(\xvec,t) \right]
\end{multlined} \label{eq:trafoHe} \\
&\bar f_i(\xvec+h\cvec_i,t+h) = f_i^*(\xvec,t) \label{eq:poststreaming} \\
&\fvec^*(\xvec,t) =
\begin{multlined}[t]
\fvec(\xvec,t) - \Lambdaop_\frac12 \left[ \fvec(\xvec,t) - \fvec^\eq(\xvec,t) \right]
\end{multlined}.
\end{align}
\end{subequations}
The difference to a direct Crank-Nicolson discretization of the
Boltzmann equation is that two explicit collisions are applied, and
not just one collision using a midpoint approximation for the
non-equilibrium distribution~\cite{Xu2003,Ubertini2010}.
%
%
To see the relation to the standard LB scheme, we note that equation
\eqref{eq:poststreaming} for the post-streaming distribution can be
rewritten as
\begin{equation}
\begin{split}
\bar f_i(\xvec+h\cvec_i,t+h) &= f_i^*(\xvec,t) \\ 
&= f_i^\eq(\xvec,t) + \sum_j (\1-\Lambdaop_\frac12)_{ij} \left[ f_j(\xvec,t) -
f_j^\eq(\xvec,t) \right] \\ 
&= f_i^\eq(\xvec,t) + \sum_j (\1-\Lambdaop_\frac12)^2_{ij} \left[ \bar f_j(\xvec,t) - f_j^\eq(\xvec,t)
\right] , \\
\end{split}
\end{equation}
which is the standard LB scheme with collision matrix
$(1-\Lambdaop_\frac12)^2$. A Chapman-Enskog expansion
\cite{ChapmanCowling} reveals that the moments of $\bar f_i$ satisfy
the Navier-Stokes equation, which allows us to obtain explicit
expressions for the transport coefficients.

The viscosity $\nu$ is given
by~\cite{McNamara1993,Dellar2001,Ladd2001,Duenweg2008}
\begin{equation}\label{eq:viscosity}
\nu = \frac{h c_s^2}{2} \frac{1+\gamma_\frac12^2}{1-\gamma_\frac12^2} 
= {h c_s^2} \left( \frac{1}{2\lambda_\frac12-\lambda_\frac12^2} - \frac{1}{2} \right)
= {h c_s^2} \left( \frac{1}{2\lambda_\frac12} - \frac{1}{4} \right) +O(h^2) ,
\end{equation}
where $\gamma_\frac12 = 1 - \lambda_\frac12$ is the eigenvalue of $( 1
- \Lambdaop_\frac12)$ corresponding to the shear modes. The relation
to the continuous Boltzmann equation with BGK collisions
\cite{Bhatnagar1954} is achieved by shifting the BGK relaxation time
$\tau$ according to the time discretization~\cite{Henon1987,Xu2003,
  Ubertini2010}, cf. Eq.~\eqref{eq:collision-operator2},
\begin{equation}\label{eq:BGKlambda}
\lambda_\frac12 = \frac{h}{2} \frac{1}{\tau + \frac{h}{4}} ,
\end{equation}
such that
\begin{equation}
\nu = {h c_s^2} \left(\frac{\tau+\frac{h}{4}}{h} -  \frac{1}{4} \right) = \tau c_s^2 ,
\end{equation}
in accordance with the continuum result \cite{ChapmanCowling}.

The viscous stress $\tens{\sigma}$ is given
by~\cite{Duenweg2008,Dellar2001}
\begin{equation}\label{eq:viscstress}
\begin{split}
\tens{\sigma} &= \frac{1}{2} \left( \1 + (\1-\Lambdaop_\frac12)^2 \right) \sum_i (\bar f_i - f^\eq_i) \cvec_i \cvec_i \\
&= \frac{(\1-\Lambdaop_\frac12)^{-1} + (\1-\Lambdaop_\frac12)}{2} \sum_i (f_i - f^\eq_i) \cvec_i \cvec_i \\
&= \sum_i (f_i - f^\eq_i) \cvec_i \cvec_i + O(h^2) .
\end{split}
\end{equation}
The second line corresponds to a midpoint scheme that averages the
stress before the last half-step collision and after the next
half-step collision. The last line shows that in the
collide-stream-collide scheme, the viscous stress is equal to the
local non-equilibrium stress up to $O(h^2)$ and can directly be
obtained from the distributions $f_i$. No additional transformation or
look-ahead collision as in the standard scheme has to be performed.

If we approximate $\mathsf{C}^\frac12 = \frac{1}{2}(\1+\mathsf{C})$,
the half-step collisions are given by \eqref{eq:half-col-approx} and
read
%
%
\begin{equation}\label{eq:half-collision}
\begin{split}
\mathsf{C}^\frac{1}{2} \fvec(\xvec,t) 
&= \fvec(\xvec,t) - \left( \1 + \frac{h}{2} \Omegaop \right)^{-1} \frac{h}{2} \Omegaop \left[ \fvec(\xvec,t) - \fvec^{\eq}(\xvec,t) \right] + O(h^2) .\\
\end{split}
\end{equation}
In this approximation, as pointed out by Dellar \cite{Dellar2013}, the
second half-step collision \eqref{eq:trafoHe} coincides with the
transformation introduced by He et al. \cite{He1998}. This
demonstrates that the different interpretations of the operator
splitting lead to equivalent LB schemes. More precisely, the LB
variables $\tilde f_i$ in \eqref{eq:LBE} correspond in fact to the
intermediate post-streaming distributions $\bar f_i$ in the splitting
scheme, and the transformation \eqref{eq:trafoHe} is necessary to
complete the full time-step.
It is to be noted that the approximation in
Eq.~\eqref{eq:half-col-approx} is insufficient for global second order
accuracy if the half-step collisions are applied in every step, since
the error is $O(h^2)$ per step. Hence, the choice $\lambda_\frac12 =
\lambda/2$ does not lead to the same viscosity as the standard scheme,
which can be easily seen from Eq.~\eqref{eq:viscosity}. The difference
can be traced back to the shift $h/4$ in Eq.~\eqref{eq:BGKlambda}
which is a consequence of the splitting of the collisions into two
half-steps. The correct relaxation parameter $\lambda_\frac12$ can
instead be directly determined from Eq.~\eqref{eq:viscosity} or
\eqref{eq:BGKlambda}. It should also be noted that it is not
straightforward to invert \eqref{eq:half-collision} if the collision
operator includes fluctuations.

The force operator $\mathsf{F}$ is readily included in the splitting
scheme $\mathsf{F}^\frac12 \mathsf{C}^\frac12 \mathsf{S}
\mathsf{C}^\frac12 \mathsf{F}^\frac12$. Equations \eqref{eq:csc} for
the intermediate steps become
\begin{subequations}
\begin{align}
&\fvec(\xvec,t) =
\begin{multlined}[t]
\fvec^\eq(\rho,\uvec(t-\frac{h}{2})) 
+ (\1-\Lambdaop_\frac12) \bigg[ \bar \fvec(\xvec,t) 
- \fvec^\eq(\rho, \uvec(t-\frac{h}{2}) \bigg] \\
+ \frac{h}{2} \vec{G}(t-\frac{h}{4})
\end{multlined} \label{eq:force-second-halfstep} \\
&\bar f_i(\xvec+h\cvec_i,t+h) = f_i^*(\xvec,t) \label{eq:force-streaming}\\
&\fvec^*(\xvec,t) = \fvec^\eq(\rho,\uvec(t)) + (\1 - \Lambdaop_\frac12) \left[ \fvec(\xvec,t) - \fvec^\eq(\rho,\uvec(t)) \right] + \frac{h}{2} \vec{G}(t+\frac{h}{4}) ,
\label{eq:force-first-halfstep}
\end{align}
\end{subequations}
where we have used midpoint forces in both forcing steps. As before,
the relation to the standard LB scheme can be elucidated in terms of
the post-streaming variables $\bar f_i(\xvec+h\cvec_i,t+h)$. Inserting
\eqref{eq:force-second-halfstep} and \eqref{eq:force-first-halfstep}
into \eqref{eq:force-streaming} we obtain 
\begin{equation}
\begin{split}
&\bar f_i(\xvec+h\cvec,t+h) \\
&= f_i^\eq(\rho,\uvec(t)) + \sum_j (\1 - \Lambdaop_\frac12)_{ij} \left[ f_j(\xvec,t) - f_j^\eq(\rho,\uvec(t)) \right] + \frac{h}{2} G_i(t+\frac{h}{4})\\
&= f_i^\eq(\rho,\uvec(t)) + \sum_j (\1 - \Lambdaop_\frac12)^2_{ij} \left[ \bar f_j(\xvec,t) - f_j^\eq(\rho,\uvec(t-\frac{h}{2})) \right] + \frac{h}{2} G_i(t+\frac{h}{4})\\
&= 
\begin{multlined}[t]
f_i^\eq(\rho,\uvec(t)) + \sum_j (\1 - \Lambdaop_\frac12)^2_{ij} \left[ \bar f_j(\xvec,t) - f_j^\eq(\rho,\uvec(t)) \right] \\
+ \sum_j (\1-\Lambdaop_\frac12)^2_{ij} \frac{h}{2} G_j(t-\frac{h}{4}) 
+ \frac{h}{2} G_i(t+\frac{h}{4}) .
\end{multlined}
\end{split}
\end{equation}
Assuming that the force term $G_i$ can be expanded around $t$, we find
up to terms of $O(h^3)$
\begin{equation}
\begin{split}
&\bar f_i(\xvec+h\cvec_i,t+h) \\
&= f_i^\eq(\rho,\uvec(t)) + 
\begin{multlined}[t]
\sum_j (\1 - \Lambdaop_\frac12)^2_{ij} \left[ \bar f_j(\xvec,t) - f_j^\eq(\rho,\uvec(t)) \right] \\
+ \sum_j \left( \1 + (\1-\Lambdaop_\frac12)^2 \right)_{ij} \frac{h}{2} G_j(t) + O(h^3)
\end{multlined} \\
&= f_i^\eq(\rho,\uvec(t)) + 
\begin{multlined}[t]
\sum_j (\1 - \Lambdaop)_{ij} \left[ \bar f_j(\xvec,t) - f_j^\eq(\rho,\uvec(t)) \right] \\
+ \sum_j (\1 - \frac{1}{2}\Lambdaop)_{ij} h G_j(t) + O(h^3) .
\end{multlined} \\
\end{split}
\label{eq:force-transformation}
\end{equation}
Equation \eqref{eq:force-transformation} contains the familiar
transformation of the force term $\vec{G}$ that is used in standard LB
schemes to maintain second-order accuracy
\cite{He1998a,Ladd2001,Guo2002,Duenweg2008}.  In the splitting scheme
it arises naturally without the need to know the transformation
a-priori.

Another aspect that is illuminated by the derivation is the
redefinition of the hydrodynamic velocity in the presence of external
forces \cite{Ladd2001,Duenweg2008}
\begin{equation}
\rho \uvec = \sum_i \bar f_i \cvec_i \cvec_i + \frac{h}{2} \Fvec ,
\end{equation}
which is simply a consequence of the second application of the force
operator $\mathsf{F}^\frac12$ after the streaming step.

Although here the primary purpose of the collide-stream-collide scheme
is to elucidate the operator splitting approach, we remark that it is
straightforward to implement using a combination of the familiar
``push'' and ``pull'' schemes \cite{Wittmann2013}. A simple
pseudo-code is listed in Algorithm~\ref{alg:csc}. If one uses a
memory-efficient data layout for intermediate results, there is only a
small overhead in form of the floating point operations for the second
relaxation step. This may be almost negligible because the performance
of the LBM on modern computing architectures is mainly limited by
memory bandwidth, and the bottleneck regarding lattice site updates
per second is really the streaming step.  Nevertheless, it may seem
that in terms of floating point operations, the collide-stream-collide
LB scheme is less efficient due to the extra collision step. On the
other hand, the standard LB scheme requires extra effort to evaluate
the viscous stress, cf. Eq.~\eqref{eq:viscstress}. As long as the
stress is only needed at the end of the calculation, the standard LB
scheme may be preferable. However, if a coupling scheme involves the
fluid stress, its computation is needed after every update. In this
case the collide-stream-collide is almost equally efficient, and it
has the additional advantage that the viscous stress is directly
available as the local non-equilibrium stress.
%

\section{Coupling methods for hybrid fluid-particle systems}

We now turn to methods for coupling immersed particles to the lattice
Boltzmann equation that allow to simulate complex fluids. Here, we
focus on objects with internal degrees of freedom, such as polymers or
vesicles. Unlike colloids, they can not easily be coupled to the fluid
by simple boundary conditions due to their complex and changing
shape. The most common methods employed in such cases are the immersed
boundary method (IBM), force coupling (FC), and external boundary
force (EBF).

\subsection{Immersed boundary method (IBM)}
\label{sec:IBM}

In the immersed boundary method \cite{Peskin2002}, a fluid-object
interface is represented by a set of Lagrangian nodes with positions
$\Xvec_i$ that are advected with the flow. Interactions on the
Lagrangian grid are applied as body forces to the fluid. The positions
of the Lagrangian nodes are additional degrees of freedom that are
updated according to
\begin{equation}
\frac{d}{d t} \Xvec_i(t) = \mathcal{P} \Xvec_i(t) = \uvec(\Xvec_i,t) ,
\end{equation}
where $\mathcal{P}$ is the propagation operator and $\uvec(\Xvec_i,t)$
is the flow velocity at the Lagrangian position $\Xvec_i$. 
The latter requires a mapping between Lagrangian positions and
Eulerian grid coordinates
\begin{equation}\label{eq:IBM-interpolation}
\uvec(\Xvec_i,t) = \Iop_a[\Xvec_i(t)] \, \uvec (\Xvec_i) = \sum_\xvec \uvec(\xvec,t) \delta_a(\Xvec_i(t)-\xvec) ,
\end{equation}
where $\delta_a$ is a discrete Dirac delta function on a
lattice with lattice spacing $a$, and $\Iop_a[\Xvec_i]$ is an
interpolation operator based on the Lagrangian positions $\Xvec_i$.
Conversely, the forces arising between the Lagrangian nodes are spread
to the grid points with the adjoint operator $\Iop_a^*$
\begin{equation}\label{eq:IBM-spreading}
\Fvec(\xvec,t) = \Iop_a^*[\Xvec_i(t)] \, \Fvec_i (\xvec) = \frac{1}{a^3} \sum_{i} \Fvec_i(t) \delta_a(\Xvec_i(t)-\xvec) .
\end{equation}
Note that $\Fvec(\xvec,t)$ is a force density while $\Fvec_i(t)$ is a
force, hence the factor $a^{-3}$ on the right hand side of
\eqref{eq:IBM-spreading}.
The construction of the discrete delta function $\delta_a$ is
described in detail in \cite{Peskin2002} and is briefly summarized in
\ref{sec:delta}.

The evolution of the complete system is described by
\begin{equation}
\frac{\d}{\d t} \{ \fvec(\xvec,t), \Xvec_i(t) \} = \left[ \mathcal{S} + \mathcal{C} + \mathcal{F} + \mathcal{P} \right] \{ \fvec(\xvec,t), \Xvec_i(t) \}.
\end{equation}
Since the interpolation operator $\Iop_a[\Xvec_i(t)]$ depends
on the positions of the Lagrangian nodes, $\mathcal{P}$ does not
commute with any of the operators $\mathcal{S}$, $\mathcal{C}$ and
$\mathcal{F}$.
Therefore, the commonly used IBM algorithm
\begin{equation}
\{ \fvec(\xvec,t+h), \Xvec_i(t+h) \} = \left[ \mathsf{P}\mathsf{S}\mathsf{C}\mathsf{F} \right] \{ \fvec(\xvec,t), \Xvec_i(t) \}
\end{equation}
is only first-order accurate. It is straightforward to write down a
formally second order scheme:
\begin{equation}\label{eq:IBM-scheme}
\{ \fvec(\xvec,t+h), \Xvec_i(t+h) \} = \left[ \mathsf{P}^\frac12 \mathsf{F}^\frac12 (\mathsf{C}^\frac12 \mathsf{S} \mathsf{C}^\frac12) \mathsf{F}^\frac12 \mathsf{P}^\frac12 \right] \{ \fvec(\xvec,t), \Xvec_i(t) \} .
\end{equation}
During the position updates $\mathsf{P}^\frac12$, the velocity is
constant but the interpolation $\Iop_a[\Xvec_i(t)]$ changes. Therefore
a simple forward Euler update is insufficient, and the Crank-Nicolson
rule becomes implicit. While one possibility is to use the midpoint
rule in each of the two half-steps $\mathsf{P}^\frac12$, we suggest
here a simpler explicit scheme that uses a predictor step for the
positions after the first half-step. Then we use a backward Euler in
the first position half-step and a forward Euler in the second half
step:
\begin{equation}
\begin{split}
\tilde \Xvec_i(t+\frac{h}{2}) &= \Xvec_i(t) + \frac{h}{2} \Iop_a[\Xvec_i(t)]\uvec(t) , \\
\Xvec_i(t+\frac{h}{2}) &= \Xvec_i(t) + \frac{h}{2} \Iop_a[\tilde\Xvec_i(t+\frac{h}{2})]\uvec(t) , \\
\Xvec_i(t+h) &= \Xvec_i(t+\frac{h}{2}) + \frac{h}{2} \Iop_a[\tilde\Xvec_i(t+\frac{h}{2})]\uvec(t+h) .
\end{split}
\end{equation}
We get an overall position update
\begin{equation}
\Xvec_i(t+h) = \Xvec_i(t) + h \Iop_a[\tilde\Xvec_i(t+\frac{h}{2})] \frac{\uvec(t) + \uvec(t+h)}{2} ,
\end{equation}
which is a trapezoidal quadrature of Eq.~\eqref{eq:IBM-interpolation}.
This scheme can be seen as the adaption of the formally second-order
method in Refs.~\cite{Lai2000,Mori2008} to the lattice Boltzmann
method. If the body force $\Fvec(\xvec,t)$ depends on the positions
$\Xvec_i$ but not on the flow velocity $\uvec$, the body forces in the
first and second half-steps $\mathsf{F}^\frac12$ are the same. The
application of the body forces thus relates to a midpoint scheme which
is interleaved with the LB collisions and streaming.
\begin{equation}
\begin{split}
\rho \uvec(t+h) &= \rho \uvec(t+\frac{h}{2}) 
+ \frac{h}{2} \Iop_a^*[\tilde\Xvec_i(t+\frac{h}{2})] \, \Fvec_i(t+\frac{h}{2}) \\
&
\begin{multlined}[t]
= \left[ \mathsf{C}^\frac12 \mathsf{S} \mathsf{C}^\frac12 \right]
\left( \rho \uvec(t) + \frac{h}{2}  \Iop_a^*[\tilde\Xvec_i(t+\frac{h}{2})] \, \Fvec_i
(t+\frac{h}{2}) \right) \\
+ \frac{h}{2} \Iop_a^*[\tilde\Xvec_i(t+\frac{h}{2})] \, \Fvec_i(t+\frac{h}{2}) . 
\end{multlined}
\end{split}
\end{equation}
The scheme in Eq.~\eqref{eq:IBM-scheme} is easily implemented by
combining a Verlet-like update for the Lagrangian nodes $\Xvec_i$ with
any second order accurate LB scheme. A pseudo-code is listed in
Algorithm \ref{alg:ibm}.

\subsection{Force coupling}
\label{sec:force-coupling}

The immersed boundary method works well if the dynamics of the fluid
and the solute objects evolve on similar time scales. In soft matter
systems, the internal motion of the solutes can be much faster than
the fluid motion and hence needs to be resolved separately. This can
be done by coupling LB to a molecular dynamics simulation where the
immersed particles evolve according to Newton's equations of motion
\cite{Ahlrichs1999}. The particle positions $\rvec_i$ are now
Lagrangian points that move with their own velocity $\vvec_i$, and the
intrinsic interactions $\Fvec_i$ enter Newton's equation of motion for
the particles.

The fluid-particle interactions are incorporated by a coupling force that
constrains the fluid flow to the motion of particles and vice-versa.
This coupling force is modeled as a drag force proportional to the
relative velocity of the particles with respect to the flow
\cite{Ahlrichs1999}
\begin{equation}\label{eq:viscous-coupling}
\begin{split}
\Fvec^{h}_i(t) &= - \gamma \left[ \vvec_i(t) - \uvec(\rvec_i,t) \right] + \zetavec_i ,
\end{split}
\end{equation}
where $\gamma$ is a friction constant, $\zetavec_i$ is a random force
required to satisfy the fluctuation-dissipation theorem, and
$\uvec(\rvec_i,t)=\Iop_a[\rvec_i(t)]\uvec(\rvec_i,t)$ is the fluid
velocity at the particle's position.
Since the coupling force must not change the total momentum of the
combined fluid-particle system, an opposite force needs to be applied
to the fluid in terms of body forces on the lattice sites within the
interpolation range around the particle
\begin{equation}
\begin{split}
\Fvec^{h}(\xvec,t) &= - \Iop_a^*[\rvec_i(t)] \Fvec_i^h(t) . 
\end{split}
\end{equation}
The interpolation and spreading operators are the same as in the
immersed boundary method. The equations of motion for the particles are
\begin{subequations}
\begin{align}
\frac{d}{d t}\rvec_i(t) &= \mathcal{P} \rvec_i(t) = \vvec_i(t) , \\
\frac{d}{d t}\vvec_i(t) &= \mathcal{F} \vvec_i(t) = \frac{1}{m_i} \left( \Fvec^{h}_i + \Fvec_i^\text{int}  \right) ,
\end{align}
\end{subequations}
where $\Fvec_i^\text{int}$ are intrinsic interactions between the
particles.

The evolution of the coupled system is now given by
\begin{equation}
\frac{\d}{\d t} \left\{ \fvec(\xvec,t), \rvec_i(t), \vvec_i(t) \right\} = \left[ \mathcal{S} + \mathcal{C} + \mathcal{F}  + \mathcal{P} \right] \left\{ \fvec(\xvec,t), \rvec_i(t), \vvec_i(t) \right\} ,
\end{equation}
where the overall force operator $\mathcal{F}$ acts simultaneously on
the particle velocities and on the fluid flow,
cf.~Eq.~\eqref{eq:LB-force-moments},
\begin{subequations}
\begin{align}
&\mathcal{F} \vvec_i(t)     = \Fvec_i^h(t) , \\ 
&\mathcal{F} \uvec(\xvec,t) = - \Iop_a^*[\rvec_i(t)] \Fvec_i^h(t) .
\end{align}
\end{subequations}
As before, we can apply the Trotter-Suzuki decomposition formula to
obtain a splitting scheme \cite{Suzuki1985}
\begin{equation}\label{eq:fc-splitting}
\left\{ \fvec, \rvec_i, \vvec_i \right\}(t+h) = \Big[ \mathsf{P}^\frac12 \mathsf{F}^\frac12 ( \mathsf{C}^\frac12 \mathsf{S} \mathsf{C}^\frac12 ) \mathsf{F}^\frac12 \mathsf{P}^\frac12 \Big] \left\{ \fvec, \rvec_i, \vvec_i \right\}(t) .
\end{equation}
This is basically a Verlet algorithm with a lattice Boltzmann update
in the middle of the velocity update.
Since the hydrodynamic coupling forces are dependent on both the
particle velocities and the fluid flow, the equations of motion are
coupled
\begin{subequations}\label{eq:fc-vel}
\begin{align}
\frac{\d}{\d t} \vvec_i(t) &= - \frac{1}{m_i} \left[ \gamma \left( \vvec_i(t) - \uvec(\rvec_i,t) \right) - \zetavec_i - \Fvec_i^\text{int} \right] , \\
\frac{\d}{\d t} \uvec(\rvec_i,t) &= \frac{1}{\rho a^3} \left[ \gamma \left( \vvec_i(t) - \uvec(\rvec_i,t) \right) - \zetavec_i \right] .
\end{align}
\end{subequations}
An accurate midpoint scheme for the velocities and the fluid flow is
\begin{subequations}\label{eq:fc-midpoint} 
\begin{align}
\vvec_i(t+h) &= \vvec_i(t) +\frac{h}{m_i} \Fvec_i^h(t+\frac{h}{2}) + \frac{h}{m_i} \Fvec_i^\text{int} , \\
\uvec(\rvec_i,t+h) &= \uvec(\rvec_i,t) - \frac{h}{\rho} \Fvec^h_i(t+\frac{h}{2})  . 
\end{align}
\end{subequations}
%
%
In Ref. \cite{Ladd2009}, a constant flow velocity $\uvec(\rvec_i,t)$
is assumed to determine the viscous coupling force. This is justified
for slowly varying flow fields that are only slightly perturbed by the
coupling, but strictly speaking it does not give an $O(h^2)$
approximation of Eq.~\eqref{eq:fc-vel}.
%
%
%
Therefore, we directly apply Crank-Nicolson discretization here
%
%
%
%
%
\begin{subequations}\label{eq:fc-cn}
\begin{align}
\begin{multlined}[t]
\frac{\vvec_i(t+h) - \vvec_i(t)}{h} = 
- \frac{\gamma}{m_i} \frac{\vvec_i(t+h) - \uvec(\rvec_i,t+h)}{2} \\
- \frac{\gamma}{m_i} \frac{\vvec_i(t) - \uvec(\rvec_i,t)}{2}
+ \frac{1}{m_i} \left( \Fvec_i^\text{int} +\zetavec_i \right) ,
\end{multlined} \\
\begin{multlined}[t]
\frac{\uvec(\rvec_i,t+h) - \uvec(\rvec_i,t)}{h} = 
\frac{\gamma}{\rho a^3} \frac{\vvec_i(t+h) - \uvec(\rvec_i,t+h)}{2} \\
+ \frac{\gamma}{\rho a^3} \frac{\vvec_i(t) - \uvec(\rvec_i,t)}{2}
- \frac{1}{\rho a^3} \zetavec_i ,
\end{multlined}
\end{align}
\end{subequations}
%
%
%
and obtain
\begin{align}
\vvec_i(t+h) &= \vvec_i(t) - \frac{\gamma^{-1}\alpha}{1+\frac{\alpha}{2}+\frac{\beta}{2}} \left[ \gamma \left( \vvec_i(t) - \uvec(\rvec_i,t) \right) - \zetavec_i - \frac{2+\beta}{2} \Fvec_i^\text{int} \right] , \\
\uvec(\rvec_i,t+h) &= \uvec_i(\rvec_i,t) + \frac{\gamma^{-1}\beta}{1+\frac{\alpha}{2}+\frac{\beta}{2}} \left[ \gamma \left( \vvec_i(t) - \uvec(\rvec_i,t) \right) - \zetavec_i + \frac{\alpha}{2} \Fvec_i^\text{int} \right] .
\end{align}
%
%
where the parameters $\alpha=\frac{\gamma h}{m_i}$ and
$\beta=\frac{\gamma h}{\rho a^3}$ were introduced.
%
Requiring consistency between \eqref{eq:fc-midpoint} and
\eqref{eq:fc-cn} we obtain the midpoint force
%
%
\begin{equation}\label{eq:modified-force}
\begin{split}
\Fvec_i^h(t+\frac{h}{2}) &= - \frac{1}{1+\frac{\alpha}{2}+\frac{\beta}{2}} \left[ \gamma \left( \vvec_i(t) - \uvec(\rvec_i,t) \right) - \zetavec_i + \frac{\alpha}{2} \Fvec_i^\text{int} \right] .
\end{split}
\end{equation}
The denominator $1+\frac{\alpha}{2}+\frac{\beta}{2}$ has the typical
form of a discrete correction, cf. Eq.~\eqref{eq:collision-operator},
and takes into account that both $\vvec_i(t)$ and $\uvec(\rvec_i,t)$
change during the time step.
%
%
%
Eq.~\eqref{eq:modified-force} is straightforward to implement and
can lead to more accurate integration of the coupling forces
\cite{Ladd2009}.
%
%
%
Note that the formulas
\eqref{eq:fc-midpoint}-\eqref{eq:modified-force} refer to the full
time step operator $\mathsf{F}$, and a half-step $\mathsf{F}^\frac12$
is achieved by substituting $h$ by $\frac{h}{2}$. For a full time step
in the splitting scheme \eqref{eq:fc-splitting}, the procedure is
applied before and after the LB update. A pseudo-code for force
coupling is listed in Algorithm~\ref{alg:fc}.

It is worthwhile to note that the viscous force coupling does not lead
to an instantaneous stick boundary condition. To see this, we
calculate the difference between the particle velocity and fluid flow
after the force application
\begin{multline}
\vvec_i(t+h) - \uvec(\rvec_i,t+h) = \frac{1 - \frac{\alpha}{2} -
\frac{\beta}{2}}{1+\frac{\alpha}{2}+\frac{\beta}{2}} \left( \vvec_i(t)
- \uvec(\rvec_i,t) \right) \\
+ \frac{\alpha+\beta}{1+\frac{\alpha}{2}+\frac{\beta}{2}}
\frac{\zetavec_i}{\gamma} +
\frac{\alpha}{1+\frac{\alpha}{2}+\frac{\beta}{2}} 
\frac{\Fvec_i^\text{int}}{\gamma} .
\end{multline}
%
%
The second term stems from the thermal fluctuations and is zero on
average. 
Requiring that the coefficient of the velocity difference vanishes
yields $\alpha+\beta = 2$ and an expression for the friction
\begin{align}\label{eq:gamma-cond}
\begin{split}
\gamma = \frac{\rho a^3}{h} \frac{2 \beta}{\alpha + \beta} = \frac{\rho a^3}{h} \frac{2}{1 + \frac{\rho a^3}{m_i}} .
\end{split}
\end{align}
In this case, the velocity difference after force application is
\begin{equation}\label{eq:slip}
\vvec_i(t+h) - \uvec(\rvec_i,t+h) = \frac{h}{2m_i} \Fvec_i^\text{int} + \frac{\zetavec_i}{\gamma} ,
\end{equation}
which does not vanish on average.
The reason for this spurious velocity slip lies in the nature of the
discretization: the velocities are effectively evaluated at the
midpoint and therefore the intrinsic forces acting during the second
half-interval are not accounted for. 
%
%
This does not cause a problem if one considers point-like particles
that are dragged through the fluid. In this case, one can tune
$\gamma$ to obtain a defined physical friction and an effective
particle size. The effective physical friction has to be re-normalized
due to the spatial interpolation on the discrete lattice and leads to
an effective size $R_\text{eff}$ of the
particles~\cite{Ahlrichs1999,Duenweg2008}
\begin{equation}
\begin{split}
\frac{1}{R_\text{eff}} &= \frac{6\pi\eta}{\gamma} + \frac{1}{g a} .
\end{split}
\end{equation}
%
%
%
%
Moreover, since the model is based on point particles, extended
objects will necessarily be under-resolved when represented by a
single particle. Therefore it is necessary to add fluctuations
$\zetavec_i$ to the coupling force in order to satisfy the fluctuation
dissipation relation \cite{Ahlrichs1999,Duenweg2008}.

In the case of more complicated objects like cells and vesicles with
specific shapes, it may me more appropriate to represent the object by
a set of surface points that are coupled to the fluid. The spurious
slip velocity \eqref{eq:slip} may then be undesirable. A remedy for
this situation is a modification of the external boundary force as
described in the following section.

\subsection{External boundary force (EBF)}
\label{sec:ebf}

The modeling of the fluid-particle interactions as a viscous force has
proved successful for simulating polymer chains in solution, where the
individual beads of the chain experience the viscous
coupling~\cite{Ahlrichs1999,Pham2009,Ladd2009}, but it may be less
appropriate for extended objects. In this case, the fluid velocity at
the surface is typically required to satisfy a no-slip boundary
condition, i.e., the fluid and the object should move with the same
velocity. Therefore, the fluid-particle interaction can also be
modeled as an external boundary force that forces the fluid to move
with the surface velocity \cite{Goldstein1993,Wu2010}. In the latter
references, the interaction force acting at a surface point is
\begin{equation}\label{eq:ebf}
\Fvec^h_i(\rvec_i,t) = - \frac{\rho a^3}{h} \left( \vvec_i(t) - \uvec(\rvec_i,t) \right) ,
\end{equation}
which is equivalent to the force coupling \eqref{eq:viscous-coupling}
with $\gamma=\frac{\rho a^3}{h}$. In this section we skip the
stochastic force $\zetavec_i$ which is straightforward to include if
demanded by the fluctuation dissipation relation for the immersed
object.

As we have seen above, however, the force in \eqref{eq:ebf} can not
lead to the no-slip boundary condition if intrinsic forces are
present.
The reason is that, in contrast to the immersed boundary method, the
intrinsic forces are applied to the particles and accelerate them
relative to the fluid. In order to enforce the no-slip boundary
condition even in the presence of intrinsic forces, we require
\begin{equation}
\begin{split}
\vvec_i(t+h) &= \uvec(\rvec_i,t+h) , \\
\vvec_i(t) + \frac{h}{m_i} \left[ \Fvec_i^h(t+\frac{h}{2}) + \Fvec_i^\text{int} \right]
&= \uvec(\rvec_i,t) - \frac{h}{\rho a^3} \Fvec_i^h(t+\frac{h}{2}) .
\end{split}
\end{equation}
Solving for the fluid-particle interaction force we get
\begin{equation}
\begin{split}
\Fvec_i^h(t+\frac{h}{2}) &= - \frac{\gamma}{\alpha + \beta} \left( \vvec_i(t) - \uvec(\rvec_i,t) \right) - \frac{\alpha}{\alpha + \beta} \Fvec_i^\text{int} .
\end{split}
\end{equation}
%
%
%
The updates for the particle and flow velocity are thus
\begin{subequations}\label{eq:modified-ebf}
\begin{align}
\vvec_i(t+h) &= \vvec_i(t) 
- \frac{\alpha}{\alpha+\beta} \left( \vvec_i(t) - \uvec(\rvec_i,t) \right) 
+ \frac{h}{m_i} \frac{\beta}{\alpha+\beta} \Fvec_i^\text{int} , \\
\uvec(\rvec_i,t) &= \uvec(\rvec_i,t)
+ \frac{\beta}{\alpha+\beta} \left( \vvec_i(t) - \uvec(\rvec_i,t) \right) 
+ \frac{h}{\rho a^3} \frac{\alpha}{\alpha+\beta} \Fvec_i^\text{int} .
\end{align}
\end{subequations}
The modified EBF in Eq.~\eqref{eq:modified-ebf} leads to an
instantaneous no-slip boundary condition at the fluid-object
interface.  The implementation is straightforward and has the same
structure as the force coupling in Algorithm \ref{alg:fc}.
%

If we introduce
\begin{align}\label{eq:immersion}
r &= \frac{\alpha}{\alpha + \beta} = \frac{1}{1 + \frac{m_i}{\rho
a^3}}, & (1-r) &= \frac{\beta}{\alpha+\beta} = \frac{1}{1 + \frac{\rho
a^3}{m_i}} ,
\end{align}
Equations~\eqref{eq:modified-ebf} can be interpreted as follows: A
fraction $r$ of the intrinsic force is included in the hydrodynamic
coupling and thus applied to the fluid, while the remaining fraction
$1-r$ is applied to the particle system. We remark that the same
equations are obtained, if we start out by assuming that the intrinsic
force is split between the fluid and the particles. That is,
Eqs.~\eqref{eq:modified-ebf} are a second order accurate scheme for
the differential equations
\begin{subequations}
\begin{align}
\frac{\d}{\d t} \vvec_i(t) &= - \frac{1}{m_i} \left[ \gamma \left( \vvec_i - \uvec(\rvec_i,t) \right) - \zetavec_i - (1-r) \Fvec_i^\text{int} \right] , \\
\frac{\d}{\d t} \uvec(\rvec_i,t) &= \frac{1}{\rho a^3} \left[ \gamma \left( \vvec_i - \uvec(\rvec_i,t) \right) - \zetavec_i  + r \Fvec_i^\text{int} \right] .
\end{align}
\end{subequations}
Equations~\eqref{eq:gamma-cond} and \eqref{eq:immersion} are then the
conditions under which the no-slip boundary condition is satisfied.

It is interesting to note that the distribution of the intrinsic force
is controlled by the ratio between the particle mass $m_i$ and
  the fluid mass within a unit cell $\rho a^3$, where $a$ is the
  lattice spacing. In the limit $m_i \gg \rho a^3$ we have $r
\rightarrow 0$, and the original EBF is recovered, where the intrinsic
force is applied to the particles only. Conversely, in the limit $m_i
\ll \rho a^3$ we have $r \rightarrow 1$ and recover the IBM where the
intrinsic force is applied to the fluid and the particles are simply
advected with the flow velocity. This shows that the immersed boundary
method corresponds to the limit where the particle mass becomes
negligible compared to the fluid mass. The dimensionless parameter $r$
controls whether the fluid is constrained to the particle velocity, or
whether the particles are constrained to the fluid flow.  We therefore
refer to $r$ as the ``immersion number''. It measures the relative
importance of inertia of the particles and the fluid.
The immersion number also depends on resolution: For fixed $\rho$ and
$m_i$, $r$ goes to zero in the continuum limit $a \rightarrow 0$.
%
%
For fixed resolution, $r$ depends only on the fraction $\alpha / \beta
= \rho a^3 / m_i$ and is independent of~$\gamma$. Hence $r$ can be
tuned to control the strength of the coupling with respect to the
transfer of intrinsic forces to the fluid. It is often desirable to
have a neutrally buoyant object, which is achieved when the immersion
number is $r=1/2$ or equivalently ${m_i}={\rho a^3}$. This choice is
also typical in other mesoscopic methods, for example, in coupling
elastic cells to multi-particle collision
dynamics~\cite{Peltomaki2013}.

The payoff for satisfying the no-slip boundary condition
instantaneously is the loss of freedom in tuning $\gamma$. The
physical friction and effective size of individual point-like
particles can not be tuned as easily any more.
However, this is fully compatible with the representation of the
solute object by a set of surface points: The object representation
has an explicit size, where the Lagrangian points are rather markers
than bead-like particles. The effective physical parameters of the
object are thus primarily determined by the boundary conditions on the
surface.

\section{Multiple time-step integration of force coupled systems}
\label{sec:multi-timestep}

Since the intrinsic forces are typically short-range, they often
require a time-step $\Delta t$ much smaller than a typical LB step
$h$. Reducing the overall time step can be costly, and as the dominant
computational effort is the fluid update, it is desirable to maintain
a large fluid time-step $h$. For accurate integration of the intrinsic
forces, we can sub-divide $h$ into an even integer number $2n$ of
sub-steps of size $\Delta t=h/(2n)$. Such a multiple time step scheme
can be written in terms of operator splitting as follows
\cite{Tuckerman1992}
\begin{equation}\label{eq:multistep}
\begin{split}
e^{h(\mathcal{S}+\mathcal{C}+\mathcal{P}+\mathcal{F})}
&\approx \left[e^{n \Delta t (\mathcal{P}+\mathcal{F})}\right] e^{h(\mathcal{S}+\mathcal{C})} \left[e^{{n\Delta t}(\mathcal{P}+\mathcal{F})}\right] \\
&= \left[e^{\Delta t(\mathcal{P}+\mathcal{F})}\right]^{n} e^{h(\mathcal{S}+\mathcal{C})} \left[e^{\Delta t(\mathcal{P}+\mathcal{F})}\right]^{n} \\
&\approx \left[ \mathsf{P}^\frac{1}{4n} \mathsf{F}^\frac{1}{2n} \mathsf{P}^\frac{1}{4n} \right]^n [\mathsf{C}^\frac{1}{2}\mathsf{S}\mathsf{C}^\frac{1}{2}] \left[ \mathsf{P}^\frac{1}{4n} \mathsf{F}^\frac{1}{2n} \mathsf{P}^\frac{1}{4n} \right]^{n} . \\
\end{split}
\end{equation}
Equation~\eqref{eq:multistep} are two Verlet-like updates, each
iterated $n$ times with a time step $\Delta t$, and arranged
symmetrically around the LB update. The symmetric arrangement is
essential for accurate operator splitting, but it seems to receive
little attention in common software implementations.

Finally, we note that it is also possible to split the update of the
full system in the following way
\begin{equation}
e^{h(\mathcal{S}+\mathcal{C}+\mathcal{P}+\mathcal{F})}
\approx  \mathsf{F}_h^\frac12 \left[ \mathsf{P}^\frac{1}{4n} \mathsf{F}_\text{int}^\frac{1}{2n} \mathsf{P}^\frac{1}{4n} \right]^n [\mathsf{C}^\frac12\mathsf{S}\mathsf{C}^\frac12] \left[ \mathsf{P}^\frac{1}{4n} \mathsf{F}_\text{int}^\frac{1}{2n} \mathsf{P}^\frac{1}{4n} \right]^n \mathsf{F}_h^\frac12,
\end{equation}
where we have decomposed the force operator into the intrinsic part
$\mathsf{F}_\text{int}$ and the coupling part $\mathsf{F}_h$. In this
arrangement, the particle system can be integrated using an ODE solver
such as \textsf{LSODE} or \textsf{VODE}
\cite{Hindmarsh1983,Brown1989}. The fluid-particle interactions are
arranged at the outside in order to satisfy the no-slip boundary
condition after a full cycle through all parts. Moreover, the number
of evaluations of the hydrodynamic coupling force is reduced by a
factor $n$. The payoff is that this may lead to inaccuracies in
tracing the flow field by the particle positions.
A detailed numerical analysis of the accuracy, efficiency and
stability of the above splitting schemes is left for future work.

\section{Conclusions}

The operator splitting approach for the lattice Boltzmann method makes
it straightforward to derive accurate time-marching schemes for
various coupling methods. The proposed second-order schemes for the
immersed boundary method, force coupling and external boundary force
improve the accuracy compared to commonly implemented first-order
algorithms. The modified external boundary force unveils the close
relation of the immersed boundary method and force coupling.  All
three methods can be cast into the form of force coupling and thus
have a unified foundation. The coupling strength can be quantified by
the ratio between fluid and particle mass. In practice, this allows us
to control the coupling by appropriate choice of the immersion number.

While finite-time accuracy of the solutions may not be the main
concern if statistical properties in equilibrium or non-equilibrium
steady states are evaluated, accurate integration methods are still
favorable to maintain long-time stability
\cite{Shardlow2003,DeFabritiis2006,Bou-Rabee2010}. Splitting methods
are fairly established in molecular dynamics simulations and lead to
superior algorithms, in particular when applied to systems with
multiple scales and long-range forces
\cite{Tuckerman1992,McLachlan2002,Bou-Rabee2013}.
It is therefore anticipated that the fluid-particle coupling methods
discussed in this work will also benefit from a systematic operator
splitting approach.
%

\section*{Acknowledgments}

I would like to thank Marisol Ripoll, Roland G. Winkler, Burkhard
D\"unweg and Anthony J. C. Ladd for valuable discussions and helpful
comments.
I also appreciate the careful and constructive review by an anonymous
referee.
Financial support from Volkswagen Stiftung under Grant No.~I/83576 is
gratefully acknowledged.


\appendix

\section{Spatial interpolation functions}
\label{sec:delta}

In the following, we outline the construction of the spatial
interpolation functions which are used in the hybrid coupling schemes
to map between the Eulerian grid and the Lagrangian particle system.
More details can be found, e.g., in \cite{Peskin2002}.

The three-dimensional delta function $\delta_a(\xvec)$ is taken
as a product of one-dimensional functions
\begin{equation}
\delta_a(\xvec) = \phi(\frac{x}{a}) \phi(\frac{y}{a}) \phi(\frac{z}{a}) ,
\end{equation}
where $x,y,z$ are the components of the vector $\xvec$. By postulate,
the function $\phi(r)$ should be continuous and satisfy the following
conditions \cite{Peskin2002}
\begin{subequations}
\begin{align}
\sum_j \phi(r-j) &= 1 \quad \forall r \label{eq:delta1} , \\
\sum_j j \phi(r-j) &= r  \quad \forall r  \label{eq:delta2} , \\
\sum_j \left( \phi(r-j) \right)^2 &= C \quad \forall r \label{eq:delta3}  ,
\end{align}
\end{subequations}
where the sum runs over all integer numbers $j$.

For reasons of computational efficiency, $\phi(r)$ should also have
bounded support. Conditions \eqref{eq:delta1} and \eqref{eq:delta2}
imply
\begin{subequations}
\begin{align}
\sum_\xvec \delta_a(\xvec - \Xvec) &= 1 , \\ 
\sum_\xvec \xvec \, \delta_a(\xvec - \Xvec) &= \Xvec ,
\end{align}
\end{subequations}
where the sum over $\xvec$ runs over all nodes of the Eulerian grid,
i.e., $\xvec = ( k a, l a, m a)$ and $k$, $l$, $m$ run over all
integer numbers.
These conditions guarantee that the total force and torque on the
Eulerian grid are the same as on the Lagrangian grid
\begin{subequations}
\begin{align}
\sum_\xvec a^3 \Fvec(\xvec,t) 
&= \sum_\xvec \sum_i \Fvec_i(t) \delta_a(\xvec - \Xvec_i) 
= \sum_i \Fvec_i(t) , \\
\sum_\xvec \xvec \times a^3 \Fvec(\xvec,t)
&= \sum_\xvec \sum_i \xvec \times \Fvec_i(t) \delta_a(\xvec - \Xvec_i)
= \sum_i \Xvec_i \times \Fvec_i(t) .
\end{align}
\end{subequations}
Moreover, \eqref{eq:delta1} and \eqref{eq:delta2} imply that smooth
functions are interpolated with second-order spatial accuracy.
The third condition \eqref{eq:delta3} can also be written as a Schwarz
inequality \cite{Peskin2002}
\begin{equation}\label{eq:delta4}
\left| \sum_j \phi(r_1-j) \phi(r_2-j) \right| \le C ,
\end{equation}
where the equal sign holds for $r_1=r_2$. This ensures that the
overlap of two Lagrangian particles is strongest when their positions
coincide, and the strength is independent of position. Condition
\eqref{eq:delta3} is thus a substitute for Galilean invariance, which
is impossible to satisfy if $\phi(r)$ has bounded support,
cf. \cite{Peskin2002}.

The three conditions \eqref{eq:delta1}-\eqref{eq:delta2} can be used to
determine a three-point interpolation function $\phi_3(r)$
\begin{equation}
\phi_3(r) = 
\begin{cases}
\frac{1}{3} \left( 1 + \sqrt{1 - 3r^2} \right) & 0 \le |r| \le \frac12 \\
\frac{1}{6} \left( 5 - 3|r| - \sqrt{6 |r| - 2 - 3r^2} \right) & \frac12 \le |r| \le \frac32 \\
0 & \frac32 \le |r| .
\end{cases}
\end{equation}
This function has a continuous derivative $\phi_3'(r)$, in contrast to
the linear interpolation used, e.g. in \cite{Ahlrichs1999}. It
therefore has the advantage that the flow gradient $\nabla \uvec$
varies smoothly and the particles do not experience jumps when
crossing grid lines. A numerical analysis of the properties of
different interpolation functions can be found in \cite{Duenweg2008}.


\clearpage

\begin{algorithm*}
\caption{Collide-Stream-Collide Lattice Boltzmann}\label{alg:csc}
\begin{algorithmic}
  \Ensure{Parallelization has communicated processor boundaries}
  \Procedure{LBUpdate}{}
    \ForAll{$(x,y,z)$}
      \State \Call{LBPush}{$x,y,z$}
      \If{$(x,y,z) \in Interior$} \Call{LBRead}{$x-1,y-1,z-1$} \EndIf
    \EndFor
    \State $t \leftarrow t+h$
  \EndProcedure
\end{algorithmic}
\hrulefill
\begin{algorithmic}
  \Require{Random numbers $r_k(x,y,z)$}
  \Procedure{LBPush}{$x,y,z$}
    \ForAll{$k$}
      \State $m_k(x,y,z) \leftarrow m_k^\eq(x,y,z) + \gamma_k m_k^\nq(x,y,z) + \sqrt{\rho}\,\varphi_k r_k(x,y,z)$ 
      \State $f_i(x+c_{ix},y+c_{iy},z+c_{iz}) \leftarrow \sum_k a^{c_i}/b_k \cdot e_{ki} m_k(x,y,z) + G_i$ 
    \EndFor
  \EndProcedure
\end{algorithmic}
\hrulefill
\begin{algorithmic}
  \Require{Random numbers $r_k(x,y,z)$}
  \Procedure{LBRead}{$x,y,z$}
    \ForAll{$k$}
      \State $m_k(x,y,z) \leftarrow \sum_i e_{ki} f_i(x,y,z)$
      \State $m_k(x,y,z) \leftarrow m_k^\eq(x,y,z) + \gamma_k m_k^\nq(x,y,z) + \sqrt{\rho}\,\varphi_k r_k(x,y,z) + \sum_i e_{ki} G_i$
    \EndFor
  \EndProcedure
\end{algorithmic}
\end{algorithm*}

\begin{algorithm*}
\caption{Immersed Boundary Method}\label{alg:ibm}
\begin{algorithmic}
  \Procedure{ImmersedBoundaryMethod}{}
    \ForAll{$i \in Particles$}
      \State $\tilde\Xvec_i \leftarrow \Xvec_i + h/2 \cdot \Iop_a[\Xvec_i] \uvec$
    \EndFor
    \State $\Fvec_i \leftarrow$ \Call{CalculateForces}{$\tilde\Xvec_i$}
    \ForAll{$i \in Particles$}
      \State $\Xvec_i \leftarrow \Xvec_i + h/2 \cdot \Iop_a[\tilde\Xvec_i] \uvec$
      \State \Call{SpreadForce}{$\Iop_a^*[\tilde\Xvec_i] \, \Fvec_i$}
    \EndFor
    \State \Call{LBUpdate}{}
    \ForAll{$i \in Particles$}
      \State \Call{SpreadForce}{$\Iop_a^*[\tilde\Xvec_i] \, \Fvec_i$}
      \State $\Xvec_i \leftarrow \Xvec_i + h/2 \cdot ( \Vvec_i + \Iop_a[\tilde\Xvec_i] \uvec )$
    \EndFor
  \EndProcedure
\end{algorithmic}
\end{algorithm*}

\begin{algorithm*}
\caption{Force Coupling}\label{alg:fc}
\begin{algorithmic}
  \Require{Random forces $\zetavec_i$}
  \Procedure{CalculateCoupling}{$\rvec_i,\vvec_i,\Fvec_i,\uvec$}
    \ForAll{$i \in Particles$}
      \State $\Fvec_i^h \leftarrow - 1/(1+\alpha/2+\beta/2) \cdot ( \gamma (\vvec_i - \Iop_a[\rvec_i] \uvec ) - \zetavec_i + \alpha/2 \cdot \Fvec_i)$
    \EndFor
  \EndProcedure
\\\hrulefill
  \Procedure{ForceCoupling}{}
    \ForAll{$i \in Particles$}
      \State $\rvec_i \leftarrow \rvec_i + \Delta t/2 \cdot \vvec_i$
    \EndFor
    \State $\Fvec_i \leftarrow$ \Call{CalculateForces}{$\rvec_i$}
    \ForAll{$i \in Particles$}
      \State $\Fvec_i^h \leftarrow$ \Call{CalculateCoupling}{$\rvec_i,\vvec_i,\Fvec_i,\uvec$}
      \State $\vvec_i \leftarrow \vvec_i + \Delta t/2 \cdot (\Fvec_i^h + \Fvec_i)$
      \State \Call{SpreadForce}{$-\Iop_a^*[\rvec_i] \, \Fvec_i^h$}
    \EndFor
    \State \Call{LBUpdate}{}
    \ForAll{$i \in Particles$}
      \State $\Fvec_i^h \leftarrow$ \Call{CalculateCoupling}{$\rvec_i,\vvec_i,\Fvec_i,\uvec$}
      \State \Call{SpreadForce}{$-\Iop_a^*[\rvec_i] \, \Fvec_i^h$}
      \State $\vvec_i \leftarrow \vvec_i + \Delta t/2 \cdot (\Fvec_i^h + \Fvec_i)$
      \State $\rvec_i \leftarrow \rvec_i + \Delta t/2 \cdot \vvec_i$
    \EndFor       
  \EndProcedure
\end{algorithmic}
\end{algorithm*}


\clearpage

\bibliographystyle{model1-num-names}
\bibliography{lb-operator-splitting}


\end{document}